# Mechanisms of temperature-dependent thermal transport in amorphous silica from machine-learning molecular dynamics

Ting Liang,[1] Penghua Ying,[2] Ke Xu,[1] Zhenqiang Ye,[3] Chao Ling,[4] Zheyong Fan,[5, *] and Jianbin Xu[1, †]

[1]*Department of Electronic Engineering and Materials Science and Technology Research Center,*
*The Chinese University of Hong Kong, Shatin, N.T., Hong Kong SAR, 999077, P. R. China*
[2]*Department of Physical Chemistry, School of Chemistry, Tel Aviv University, Tel Aviv, 6997801, Israel*
[3]*College of Materials Science and Engineering, Shenzhen University, Shenzhen 518055, China*
[4]*School of Science, Harbin Institute of Technology, Shenzhen, 518055, P. R. China*
[5]*College of Physical Science and Technology, Bohai University, Jinzhou 121013, P. R. China*
(Dated: November 1, 2023)

Amorphous silica (a-SiO$_2$) is a foundational disordered material for which the thermal transport properties are important for various applications. To accurately model the interatomic interactions in classical molecular dynamics (MD) simulations of thermal transport in a-SiO$_2$, we herein develop an accurate yet highly efficient machine-learned potential model that allowed us to generate a-SiO$_2$ samples closely resembling experimentally produced ones. Using the homogeneous nonequilibrium MD method and a proper quantum-statistical correction to the classical MD results, quantitative agreement with experiments is achieved for the thermal conductivities of bulk and 190 nm-thick a-SiO$_2$ films over a wide range of temperatures. To interrogate the thermal vibrations at different temperatures, we calculated the current correlation functions corresponding to the transverse acoustic (TA) and longitudinal acoustic (LA) collective vibrations. The results reveal that below the Ioffe-Regel crossover frequency, phonons as well-defined excitations, remain applicable in a-SiO$_2$ and play a predominant role at low temperatures, resulting in a temperature-dependent increase in thermal conductivity. In the high-temperature region, more phonons are excited, accompanied by a more intense liquid-like diffusion event. We attribute the temperature-independent thermal conductivity in the high-temperature range of a-SiO$_2$ to the collaborative involvement of excited phonon scattering and liquid-like diffusion in heat conduction. These findings provide physical insights into the thermal transport of a-SiO$_2$ and are expected to be applied to a vast range of amorphous materials.

## I. INTRODUCTION

Amorphous silica (a-SiO$_2$), owing to its excellent thermal stability and insulating properties, plays a pivotal role in the electronics and semiconductor industries, such as serving as a passivation layer in semiconductor chips [1] and a charge storage layer in metal oxide memory devices [2]. As the characteristic dimensions of integrated circuits shrink into the scale comparable to the mean free paths (MFPs) of heat carriers [3], understanding the thermal transport properties of bulk and thin-film a-SiO$_2$ at the atomic level is highly essential for enhancing device performance and lifespan. Unlike the well-defined phonons in crystalline silica, it is widely believed that the notion of phonons becomes ill-defined in a-SiO$_2$ due to the complexity and the loss of the long-range lattice periodicity in amorphous systems [4]. The phonon-gas model is accordingly failing [5–8], and there are ongoing debates about the physical picture of thermal vibration and transport in amorphous systems [9, 10].

Over the past few decades, numerous theoretical frameworks and computational methods have been proposed to comprehend and explore the thermal vibration and transport characteristics of amorphous materials. By decomposing the atomic vibrations into individual normal modes, Allen and Feldman [5–7, 11] suggested categorizing these quasi-particles into three types: propagons, diffusons, and locons, corresponding to low, medium, and high-frequency vibrations, respectively. As the Allen-Feldman (AF) formulation [11] neglects anharmonicity [12], Isaeva *et al.* [13] have recently developed the quasi-harmonic Green-Kubo (QHGK) model to investigate thermal transport in amorphous materials, allowing for the simultaneous consideration of anharmonicity and disorder. Combining the Peierls-Boltzmann equation [14, 15] describing the particle-like propagation and the AF formulation, the recently developed Wigner transport equation (WTE) [16, 17] establishes a unified formalism for microscopic heat conduction in both crystalline and disordered systems and has been applied to study thermal transport in a-SiO$_2$ [18, 19]. It is noteworthy that the QHGK and WTE methods, though originating from distinct perspectives, have been formally shown to be equivalent approaches for thermal transport in solids and can provide the same levels of approximation [20, 21].

The above frameworks provide comprehensive insights into thermal transport in amorphous materials, but their computational cost scales unfavorably with respect to the number of atoms and they are only applicable to relatively small periodic cells. Recently, based on hydrodynamic arguments, Fiorentino *et al.* [22] demonstrated that the bulk thermal conductivity of glasses can be extrapolated from medium-sized finite models. However, hydrodynamic extrapolation is still an approximate method in nature, and its applicability to various amorphous systems still requires extensive validation. Molecular dynamics (MD) is another popular method for studying thermal transport in amorphous materials [23–30] due to its natural incorporation of the full anharmonicity and the linear-scaling computational cost with respect to the simulation cell size. Particularly, it is also the best approach for generating numerical a-SiO$_2$ samples. However, an important prerequisite for this purpose is to have a reliable model for describing the inter-atomic interactions. It has been demonstrated that the Gaussian approximation potential (GAP), which is a machine-learned potential (MLP), can achieve higher accuracy for many physical properties than empirical ones for a-SiO$_2$ [31]. Motivated by this, we here construct another MLP for a-SiO$_2$ based on the neuroevolution potential (NEP) approach [32–34], reusing the training data for the GAP model [35]. The NEP model can achieve an accuracy compared to the GAP model, but with significantly higher computational efficiency, which allows for a comprehensive study of the structural and thermal transport properties of a-SiO$_2$ using large-scale MD simulations. With a low quenching rate of $10^{11}$ K s$^{-1}$ and a relatively large simulation cell containing 73 728 atoms, the generated a-SiO$_2$ samples exhibit structural characteristics that are in close agreement with experiments. Using the efficient homogeneous nonequilibrium molecular dynamics (HNEMD) method and the established quantum-statistical correction for disordered systems [29, 36], our predicted thermal conductivity agrees well with experiments for both bulk [37, 38] and thin-film [39] a-SiO$_2$ in a wide range of temperatures.

---

* brucenju@gmail.com
† jbxu@ee.cuhk.edu.hk

Further, similar to liquids [40] or liquid-like matters [41, 42], we endeavor to comprehend the thermal vibrations of a-SiO$_2$ at various temperatures by considering collective excitations, which have been experimentally observed and proposed to understand thermal transport in amorphous silicon system [43, 44]. The presence of long-range disorder and atomic instability, similar to liquids, makes collective excitations a reasonable description of the thermal vibrations in amorphous materials. Utilizing particle velocities and trajectories obtained from high-accuracy MD simulations, we calculated the transverse and longitudinal current correlation functions at different temperatures, corresponding to transverse acoustic (TA) and longitudinal acoustic (LA) collective vibrations [44–46], respectively. Within the considered temperature range, both TA and LA vibrations exhibit pronounced lattice dispersion below the Ioffe-Regel crossover frequency [44, 47, 48], revealing that the phonons are well-defined excitations, whereas above the Ioffe-Regel crossover frequency, the dispersion is absent and exhibits a liquid-like overdamping decay behavior [40–42, 49, 50]. More specifically, at low temperatures, the phonon vibrations dominate, and the thermal conductivity of the a-SiO$_2$ rises with increasing temperature. At high-temperature regions, the intense phonon scattering and the liquid-like diffusion [47, 49–51] collaboratively contribute to heat conduction, leading to the temperature-insensitive thermal conductivity observed in a-SiO$_2$.

## II. NEP MODEL TRAINING FOR A-SIO$_2$

### A. The NEP formalism

The NEP approach [32–34] is currently a popular MLP widely used in thermal transport studies [29, 36, 52–55]. It is based on a neural network (NN) and is trained using the separable natural evolution strategy (SNES) [56]. Following the standard Behler-Parrinello high-dimensional NN potential approach [57], the site energy of atom $i$ is taken as a function of a descriptor vector with $N_\text{des}$ components, $U_i(\mathbf{q}) = U_i\left(\{q_\nu^i\}_{\nu=1}^{N_\text{des}}\right)$. A feed-forward NN with a single hidden layer with $N_\text{neu}$ neurons is used to represent the site energy $U_i$:

$$U_i = \sum_{\mu=1}^{N_\text{neu}} w_\mu^{(1)} \tanh\left(\sum_{\nu=1}^{N_\text{des}} w_{\mu\nu}^{(0)} q_\nu^i - b_\mu^{(0)}\right) - b^{(1)}, \quad (1)$$

where $\tanh(x)$ is the activation function, $\mathbf{w}^{(0)}$, $\mathbf{w}^{(1)}$, $\mathbf{b}^{(0)}$, and $b^{(1)}$ are the trainable weight and bias parameters in the NN. In the NEP model, the local atom-environment descriptor $q_\nu^i$ is formed by a number of radial and angular components. The radial descriptor components, which are determined solely by atom-pair distances, are constructed based on Chebyshev polynomials, while the angular descriptor components, which also depend on angular information, are constructed based on spherical harmonics analogous to the atomic cluster expansion approach [58]. For more detailed descriptions of the NEP model, the reader is referred to the literature [32–34].

### B. The training and test data

We re-used the training data set constructed by Erhard *et. al* [31], but with some refinement. The original data set comprises 3074 structures, including 2000 crystalline ones, 939 amorphous or liquid ones, and 135 dimer or cluster ones [35]. Each structure contains reference energy, force, and virial from density-functional theory (DFT) calculations with the SCAN functional [59]. This data set has been used to construct a GAP model for SiO$_2$ [31]. Through comprehensive benchmark, the GAP model [31] has been demonstrated to outperform many empirical potentials in characterizing various properties of silica, as it incorporates a diverse set of configurations, including crystalline, liquid, and particularly a-SiO$_2$ structures.

During a preliminary training of the NEP model using this original data set, we identified a large number of outliers. We do not know the exact cause for the outliers, but empirically found that much better training accuracy for the remaining structures can be obtained by removing those outliers from the training data set. By doing so, we obtained a refined data set containing 2609 structures, including 1832 crystal ones, 761 amorphous or liquid ones, and 16 dimer or cluster ones. We then randomly selected 2348 structures (205760 atoms in total) for training, and used the remaining 261 ones (23135 atoms in total) for testing.

## C. Training results

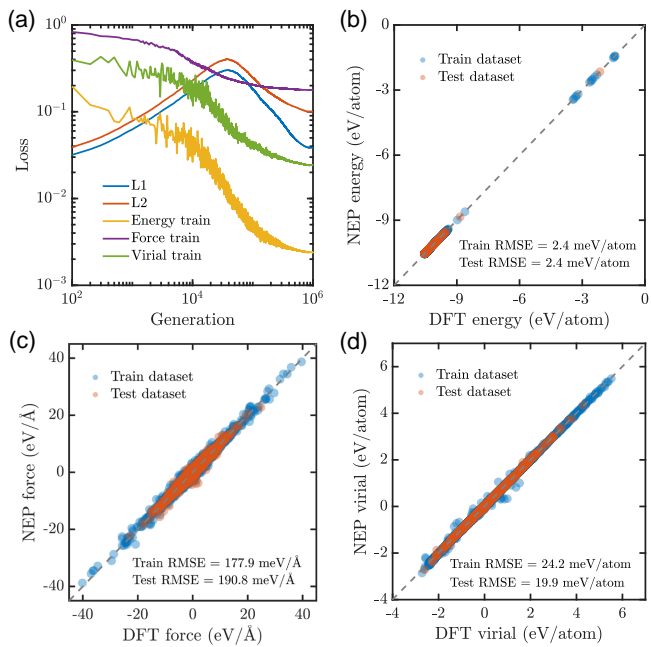

FIG. 1. (a) Evolution of the various terms in the loss function for the training data set with respect to the generation, including the $\mathcal{L}_1$ and $\mathcal{L}_2$ regularization, the energy root-mean-square error (RMSE), the force RMSE, and the virial RMSE. (b)-(d) The comparison between the NEP predictions and DFT reference data of energy, force, and virial for the training and testing data sets. The dashed lines in (b)-(d) are a guide for the eyes.

Using the refined training data set, we trained a NEP model (see Supplemental Note S1 for details on the hyperparameters) for a-$SiO_2$ using the GPUMD package [60]. The evolution of the training accuracy is shown in Fig. 1 (a). The parity plots of energy, force, and virial shown in Fig. 1 (b)-(d) demonstrate the high accuracy of the NEP model. Quantitatively, the root mean square errors (RMSEs) for energy, force, and virial are 2.4 meV/atom, 177.9 meV/Å, and 24.2 meV/atom in the training data set and are 2.4 meV/atom, 190.8 meV/Å, and 19.9 meV/atom in the test data set. The accuracy is comparable to that achieved by the GAP model [31] (see Fig. S1).

To demonstrate the generality of the NEP model, we computed the phonon dispersion for three silica crystal structures (see Fig. S2), showing good agreement with the experimental results. Our NEP model can achieve a computational speed of $1.1 \times 10^7$ atom-step per second in MD simulations, using a single Nvidia GeForce RTX 4090 GPU card. The good speed and accuracy of the NEP model provide foundations for a comprehensive study of the heat transport in $SiO_2$.

## III. SAMPLE GENERATION AND STRUCTURAL VALIDATION

### A. Generating a-$SiO_2$ samples

A rational structure determines a reliable physical property. Here, the classical MD simulations with a sophisticated melt-quench-anneal process are employed to prepare reasonable samples of a-$SiO_2$. All the MD simulations are performed using the GPUMD package [60] (version 3.6). Taking a silica crystal from Material Project [61] in the cubic $Im-3\overline{m}$ space group as the initial structure, it is randomized at 5000 K within the $NVT$ (constant number of atoms $N$, constant volume $V$, and constant target temperature $T$) ensemble, then hold at 4000 K to generate a melt. During this stage, we tested various melting temperatures (see Fig. S4) and found that 4000 K is the most suitable for generating the final a-$SiO_2$ structure. Following a supercooled process with target temperature decreases linearly from 4000 K to a final temperature $T$ (ranging from 20 to 2000 K) and a given quenching rate (reducing from $5 \times 10^{12}$ K s$^{-1}$ to $1 \times 10^{11}$ K s$^{-1}$), we produce the initial a-$SiO_2$ structure under the $NpT$ ensemble with zero target pressure $p$. Subsequently, the annealing process for 1 ns is employed to eliminate internal stresses within the system and obtain the final a-$SiO_2$ structure, as shown in Fig. 2 (a). During the melt-quench-anneal process, the volume of the simulated system steadily decreases (see Fig. S3 for energy and volume evolution), suggesting that the slow quenching rate drives the simulated system transition from a low-density liquid to a high-density amorphous state.

Further, we have visualized the evolution of the coordination number (CN) of the model during the melt-quench-anneal protocol, as depicted in Fig. 2 (b). The initial crystalline structure evolves gradually from point **A** (corresponding to randomization of the liquid structure), which is a mixture of high and low CNs, to point **E** (corresponding to solid a-$SiO_2$), which overwhelmingly contains the two-fold and four-fold coordination environments. In a plausible a-$SiO_2$ structure, oxygen atoms are coordinated by two silicon atoms, while silicon atoms are coordinated by four oxygen atoms [31, 62], consistent with the structures of the ambient pressure crystalline polymorphs. The correctly coordinated atoms demon-strate the NEP model can significantly prevent the number of defects in the a-SiO$_2$ structures. Fig. 2 (c) provides a statistical overview of the evolution of different CNs during the melt-quench-anneal process, highlighting that the trained NEP model is capable of generating highly plausible configurations for a-SiO$_2$.

### B. Validating a-SiO$_2$ samples

To characterize the short-range ordered bond motifs for the generated a-SiO$_2$ structure, we computed the pair-correlation function $g(r)$ at 300 K and a quenching rate of $10^{11}$ K s$^{-1}$. As shown in Fig. 3 (a), the NEP calculations for the first and second peaks of $g(r)$ align remarkably well with experimental measurements, both in terms of peak positions and heights. Additionally, the medium-range ordering, a typical characteristic of amorphous structures [4, 64], is often captured experimentally by the static structure factor $S(q)$ of X-ray diffraction. As shown in Fig. 3 (b), the computed X-ray $S(q)$ qualitatively reproduces the experimental features, especially for the height and position of the second diffraction peak. Precisely, reflecting the second-neighbors distances with respect to the second peak of $g(r)$ [29, 65], the position of the first diffraction peak of $S(q)$ coincides well with the experimental results. However, the NEP model conspicuously underestimates the height on the first diffraction peak, even at a low quenching rate of $10^{11}$ K s$^{-1}$. This is presumably because, although the quenching rate is slower than in quantum-mechanical simulations, it remains much faster than in most experimental settings [66]. The calculations for $g(r)$ and $S(q)$ are both performed using the ISAACS package [67], with the utilization of its embedded smoothing functionality in the case of $g(r)$.

We have compared the $g(r)$ and $S(q)$ calculated for different system sizes at the quench rate of $10^{11}$ K s$^{-1}$ with the experimental results, as shown in Fig. S5. There is no notable variation observed in the first and second peaks of $g(r)$ as the system size increases, suggesting that the employed system size is capable of encompassing short-range structural information. Concerning the $S(q)$, its first diffraction peak exhibits a gradual rise as the system increases. Moreover, the $g(r)$ and $S(q)$ calculations at various quenching rates by the NEP model (refer to Fig. S6) suggest that lower quenching rates produce samples that are closer to experimental outcomes. Striking a balance between computational efficiency and accuracy, we employed a system consisting of 73,728 atoms and a $10^{11}$ K s$^{-1}$ quenching rate for subsequent thermal conductivity calculations.

## IV. HEAT TRANSPORT PROPERTIES OF A-SIO$_2$

### A. Thermal conductivity of a-SiO$_2$ at various temperatures

We use the efficient HNEMD method [73] to calculate the thermal conductivity of a-SiO$_2$. In this method, an external driving force

$$\boldsymbol{F}_i^{\text{ext}} = E_i \boldsymbol{F}_e + \boldsymbol{F}_e \cdot \mathbf{W}_i, \quad (2)$$

is exerted on each atom $i$, driving the system out of equilibrium. Here, $E_i$ is the total energy of atom $i$, while $\boldsymbol{F}_e$ is the driving force parameter with the dimension of inverse length, and $\mathbf{W}_i$ is the virial tensor of atom $i$,

$$\mathbf{W}_i = \sum_{j \neq i} \left( \frac{\partial U_j}{\partial \boldsymbol{r}_{ji}} \otimes \boldsymbol{r}_{ij} \right). \quad (3)$$

Here, $U_i$ is the potential energy of atom $i$, and $\boldsymbol{r}_{ij} \equiv \boldsymbol{r}_j - \boldsymbol{r}_i$, $\boldsymbol{r}_i$ being the position of atom $i$. The driving force will induce an ensemble averaged (represented by $\langle \rangle$) steady-state nonequilibrium heat current $\boldsymbol{J}$ (for heat current for many-body potentials, see Ref. 74)

$$\frac{\langle J^\alpha \rangle}{TV} = \sum_\beta \kappa^{\alpha\beta} F_e^\beta, \quad (4)$$

where $\kappa^{\alpha\beta}$ is the $\alpha\beta$-component of the thermal conductivity tensor, $T$ is the system temperature, and $V$ is the system volume. For the employed a-SiO$_2$ samples, which possess cubic symmetry and are isotropic, we can consider any direction, e.g., the $z$ direction, in which case we have

$$\kappa^{zz} = \frac{\langle J^z \rangle}{TVF_e^z}. \quad (5)$$

Since the heat current $J^z$ is proportional to the magnitude of the driving force parameter $F_e^z$, cautious selection of $F_e^z$ is essential to secure a large signal-to-noise ratio within the linear-response regime of the systems. In this work, the value of $F_e^z = 2 \times 10^{-4}$ Å$^{-1}$ was tested to be appropriate for all the a-SiO$_2$ samples. For each sample, we performed five independent simulations and calculated a suitable estimate of statistical error.

Employing the HNEMD method, we conducted preliminary benchmark testing to assess the influence of various factors on the thermal conductivity $\kappa$. Referring to Fig. S7, the thermal conductivity demonstrates insensitivity to the quenching initiation temperature (as indicated by point **B** in Fig. 2 (a)) above 4000 K. Despite HNEMD being regarded as a method akin to Green-Kubo method [75, 76] for estimating the thermal conductivity of infinite systems (bulk systems), it still exhibits finite-size effects, which arise from the truncation of certain long-wavelength vibrations by a finite simulation unit and the ignoring of some scattering events [77]. Consequently, we evaluate the impact of simulation size on the thermal conductivity of a-SiO$_2$ in Fig. S8. The findings indicate that the thermal conductivities of the diverse-sized a-SiO$_2$ systems remain within the range of statistical error. Additionally, the thermal conductivity $\kappa$ of a-SiO$_2$ demonstrates a slight rise with the decrease in quenching rate (see Fig. S9), suggesting that lower quenching rates produce a more ordered structure with higher thermal conductivity. As mentioned in Sect. III, opting for a system of 73,728 atoms and a quenching rate of $10^{11}$ K s$^{-1}$ for subsequent thermal conductivity calculations is a computationally feasible and secure choice.

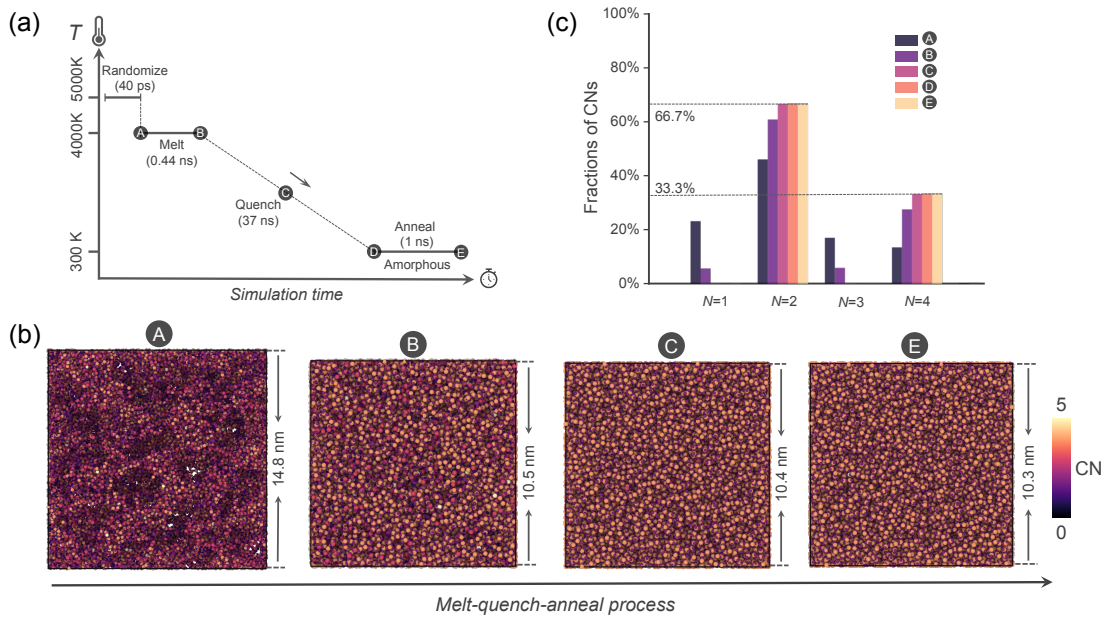

FIG. 2. Visualization of a melt-quench-anneal process for generating a-SiO$_2$ structures. (a) The temperature protocol during the structure generation process. Here, the target temperature is set to 300 K, and the quenching process lasts for 37 ns at a rate of $10^{11}$ K s$^{-1}$. (b) Structural snapshots during the melt-quench-anneal process, including point **A** after structural randomization, point **B** after melting, point **C** during quenching, and point **E** after annealing. Coordination numbers (CN, spatial cut-off = 2.0 Å), are indicated by colour coding. All structural snapshots were created using the OVITO package [63]. (c) Evolution of fractions with different CNs during the melt-quench-anneal process.

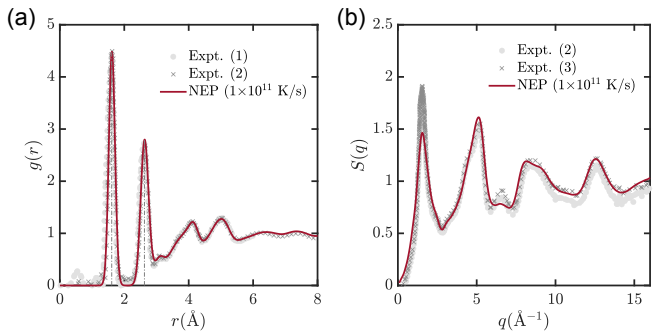

FIG. 3. The comparison of (a) pair-correlation function $g(r)$ and (b) X-ray structure factor $S(q)$ of a-SiO$_2$ calculated by NEP model with experimental measurements ("1"; ref. [68], "2"; ref. [69], and "3"; ref. [70]). At the same temperature of 300 K as in the experiments, the $g(r)$ and $S(q)$ by NEP model are obtained from a 73,728-atom system under a quenching rate of $10^{11}$ K s$^{-1}$. The vertical dash lines in (a) mark the experimental [71] first peak of Si-O at 1.61 Å and second peak of O-O at 2.63 Å, respectively.

Fig. 4 (a) compares the HNEMD-calculated thermal conductivity $\kappa$ of a-SiO$_2$ at different temperatures with experimental ones. The outcomes from HNEMD (depicted as squares) align with experimental ones primarily around room temperature, showing slight deviation at high temperatures and marked overestimation at low temperatures. In precise terms, the thermal conductivity yielded by HNEMD calculation and experimental measurement [37] is approximately 1.44 and 1.40 W m$^{-1}$ K$^{-1}$, respectively, at a temperature of 350 K. Of interest, at high temperatures, the experimentally measured thermal conductivity [72] rises with increasing temperature, which could be owing to the contribution of radiative energy to heat transfer [78–80]. Thus, in Fig. 4 (a), the discrepancy between theory and experiments at high temperatures might be due to the non-negligible radiative contributions in the experiments. Note that classical MD simulations follow the classical Boltzmann statistics, whereas at low temperatures (especially below the Debye temperature, about 495 K for silica [81, 82]), quantum Bose-Einstein vibrational statistics play a pivotal role in determining the 'correct' thermal conductivity. Regarding the pronounced overestimation at low temperatures, we attribute this primarily to the inability of MD simula-tions to account for quantum statistics. Therefore, it becomes necessary to employ a feasible quantum-correction method to rectify the thermal conductivity $\kappa$ at low temperatures.

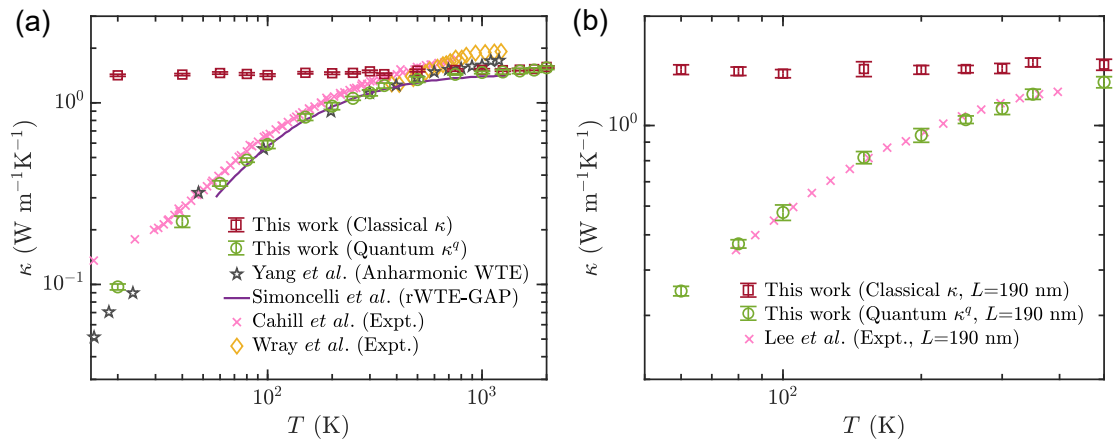

FIG. 4. Thermal conductivity $\kappa$ of a-SiO$_2$ acquired via various techniques at different temperatures. (a) Comparison of HNEMD-calculated thermal conductivity $\kappa$ of a-SiO$_2$ in the bulk limit with respect to other theoretical and experimental results. Theoretical results include those obtained by Yang et al. [18] using the anharmonic WTE, as well as those derived by Simoncelli et al. [19] using the regularized WTE (with the GAP model [31] as input, so denoted as rWTE-GAP). Experimental measurements extracted from Cahill et al. [37, 38] and Wray et al. [72]. (b) Comparison of the $\kappa$ of a-SiO$_2$ samples with a thickness of $L = 190$ nm at different temperatures to experimental results [39]. The sample thickness on the thermal transport direction is indicated as $L$.

### B. Quantum-statistical correction

Within the framework of the HNEMD approach [73], there is a feasible quantum correction method based on spectral thermal conductivity for disordered systems and has been successfully used for amorphous silicon [29], amorphous HfO$_2$ [30], and liquid water [36]. For the spectrally decomposed thermal conductivity, one can first calculate the virial-velocity correlation function

$$\boldsymbol{K}(t) = \sum_i \langle \mathbf{W}_i(0) \cdot \boldsymbol{v}_i(t) \rangle, \qquad (6)$$

where $\boldsymbol{v}_i$ is the velocity of atom $i$. The spectral thermal conductivity is then derived as

$$\kappa^{zz}(\omega, T) = \frac{2}{VTF_e^z} \int_{-\infty}^{\infty} dt e^{i\omega t} K^z(t). \qquad (7)$$

This $\kappa^{zz}(\omega, T)$ is classical, and the total thermal conductivity $\kappa^{zz}(T)$ is then obtained as an integral over the entire frequency range as

$$\kappa^{zz}(T) = \int_0^{\infty} \frac{d\omega}{2\pi} \kappa^{zz}(\omega, T). \qquad (8)$$

With the classical spectral thermal conductivity available, then a quantum-corrected spectral thermal conductivity $\kappa^q(\omega, T)$ can be obtained by multiplying $\kappa^{zz}(\omega, T)$ with a probability $p(x)$ of activation frequency between quantum and classical modal heat capacity [24, 29, 30, 36, 83]

$$\kappa^q(\omega, T) = \kappa^{zz}(\omega, T) p(x), \qquad (9)$$

where

$$p(x) = \frac{x^2 e^x}{(e^x - 1)^2}. \qquad (10)$$

Here, $x = \hbar\omega/k_B T$, $\hbar$ is the reduced Planck constant, and $k_B$ is the Boltzmann constant. The total quantum-corrected thermal conductivity can be straightforwardly represented as

$$\kappa^q(T) = \int_0^{\infty} \frac{d\omega}{2\pi} \kappa^q(\omega, T). \qquad (11)$$

Fig. 5 presents the spectral thermal conductivity with classical and quantum correction at various temperatures. At low temperatures, the primary contribution of quantum-corrected spectral thermal conductivity arises from low-frequency vibrational modes ($\omega/2\pi < 10$ THz), with the complete suppression of high-frequency vibrational modes, indicating a substantial impact of quantum correction. This is because, in classical MD simulations, the vibrations within the disordered system follow the classical Boltzmann statistics, resulting in complete excitation of all vibrational modes irrespective of temperature and frequency. In reality, however, high-frequency vibrational modes should be frozen at low temperatures according to the quantum Bose-Einstein statistics [29, 36, 84]. Fig. S10 (a) presents a comparison between the normalized vibrational density of states at 300 K and experimental data [85], clearly revealing a qualitative agreement.

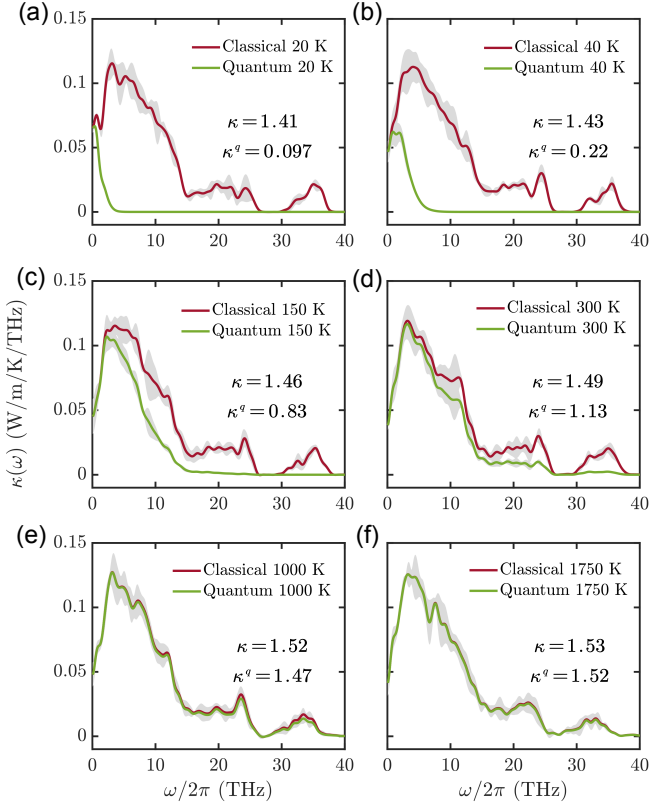

FIG. 5. Classical and quantum-corrected spectral thermal conductivity of a-SiO$_2$ at temperatures (a) 20, (b) 40, (c) 150, (d) 300, (e) 1000, and (f) 1750 K. The gray shading represents the standard deviation obtained from five independent simulations. The label on each panel represents the total thermal conductivity value obtained by integrating the entire spectrum.

The quantum-corrected HNEMD results (denoted by circles in Fig. 4 (a)) agree well with the experimental ones by Cahill *et al.* [37, 38] at low temperatures, although there is a minor underestimation at 20 K. We have further presented the results from other theoretical predictions, including those obtained by Yang *et al.* [18] using the anharmonic WTE, as well as those derived by Simoncelli *et al.* [19] using the regularized WTE. All the methods lead to satisfactory agreement with experiments.

### C. Length dependence of thermal conductivity for a-SiO$_2$

Akin to amorphous silicon [86, 87], a-SiO$_2$ thin films exhibit strong size effects for thermal conductivity at nanoscale. To verify the efficacy of the aforementioned quantum corrections on the a-SiO$_2$ thin films, we evaluated the quantum-corrected thermal conductivity for a film thickness of $L = 190$ nm, a dimension for which experimental data [39] are accessible. Leveraging the HNEMD approach in conjunction with the nonequilibrium molecular dynamics (NEMD) simulations can achieve this purpose efficiently and elegantly. In this framework, the spectral conductance in the ballistic regime (low $T$ and short $L$) is first calculated using NEMD simulations, and then the thermal conductance $G(\omega)$ (see Fig. S10 (b)) can also be spectrally decomposed:

$$G^{zz}(\omega) = \frac{2}{V \Delta T} \int_{-\infty}^{\infty} \mathrm{d}t e^{i\omega t} K^z(t), \qquad (12)$$

where $\Delta T$ is the temperature difference between the heat source and heat sink in the NEMD setup. Subsequently, one can calculate the spectral-decomposed MFP as $\lambda^{zz}(\omega, T) = \kappa^{zz}(\omega, T)/G^{zz}(\omega)$ (see Fig. S10 (c)). Within the ballistic transport NEMD simulation, the temperature is designated as 20 K, and the effective thickness of the a-SiO$_2$ system is about 2.4 nm. With the spectral MFP, we can obtain the quantum-corrected thickness-dependent thermal conductivity at a desired temperature (see Fig. S10 (d))

$$\kappa^{\mathrm{q}}(L, T) = \int \frac{\mathrm{d}\omega}{2\pi} \kappa^{\mathrm{q}}(L, \omega), \qquad (13)$$

where

$$\kappa^{q}(L, \omega) = \frac{\kappa^{q}(\omega, T)}{1 + \lambda^{zz}(\omega, T)/L}. \qquad (14)$$

At the same film thickness of $L = 190$ nm as in the experiments [39], our quantum-corrected thickness-dependent thermal conductivity results exhibit remarkably good agreement with the experimental data from Lee *et al.* [39] (within the temperature range of 80 to 400 K), as depicted in Fig. 4 (b). Such an agreement effectively substantiates the effectiveness of our quantum correction approach. In Fig. S11, we provide the corrected thermal conductivities for a-SiO$_2$ films of 190 nm thickness spanning the temperature range of 40 to 2000 K for comparative analysis.

### D. Origins of temperature dependence

The current correlation functions, which can be considered as a spatially dependent generalization of the velocity correlation function [45] and are closely related to the phonon spectral energy density [54, 88], are employed to examine the behavior of collective excitations [44–46] in a-SiO$_2$ at varying temperatures. Analogous to the density of atoms

$$n(\mathbf{r}, t) = \sum_i^N \delta \left( \mathbf{r} - \mathbf{r}_i(t) \right), \qquad (15)$$

one can consider the current density $\boldsymbol{j}(\boldsymbol{r}, t)$ of a disordered system based on particle velocities [40, 41, 45]. The

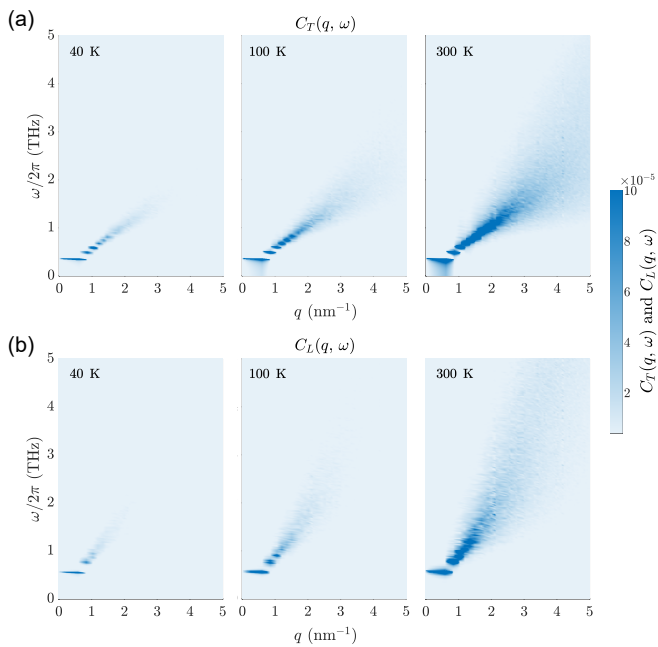

FIG. 6. (a) Transverse and (b) longitudinal current correlation functions as a function of $\boldsymbol{q}$ vector and $\omega/2\pi$ from 40 to 300 K.

current density $\boldsymbol{j}(\boldsymbol{r},t)$ is given by

$$\boldsymbol{j}(\boldsymbol{r},t) = \sum_i^N \boldsymbol{v}_i(t)\delta\left(\boldsymbol{r} - \boldsymbol{r}_i(t)\right), \quad (16)$$

with the Fourier components $\boldsymbol{j}(\boldsymbol{q},t)$

$$\boldsymbol{j}(\boldsymbol{q},t) = \sum_i^N \boldsymbol{v}_i(t)e^{i\boldsymbol{q}\cdot\boldsymbol{r}_i(t)}, \quad (17)$$

where $N$ is the number of atoms, $\boldsymbol{v}_i(t)$ is the velocity of atom $i$ at time $t$, and $\boldsymbol{r}_i(t)$ is the position of atom $i$ at time $t$. Here, the current density is a vector that can be decomposed into a transverse part $\boldsymbol{j}_T(\boldsymbol{q},t)$ perpendicular to the $\boldsymbol{q}$-vector and a longitudinal part $\boldsymbol{j}_L(\boldsymbol{q},t)$ containing the parallel component

$$\boldsymbol{j}_T(\boldsymbol{q},t) = \sum_i^N \left[\boldsymbol{v}_i(t) - (\boldsymbol{v}_i(t)\cdot\hat{\boldsymbol{q}})\hat{\boldsymbol{q}}\right]e^{i\boldsymbol{q}\cdot\boldsymbol{r}_i(t)}, \quad (18)$$

$$\boldsymbol{j}_L(\boldsymbol{q},t) = \sum_i^N \left(\boldsymbol{v}_i(t)\cdot\hat{\boldsymbol{q}}\right)\hat{\boldsymbol{q}}e^{i\boldsymbol{q}\cdot\boldsymbol{r}_i(t)}, \quad (19)$$

where $\hat{\boldsymbol{q}}$ denotes the unit vector. Analogous to the intermediate scattering function (see Ref. [45] for details), we can calculate the current correlation functions as

$$C_T(\boldsymbol{q},\omega) = \int_{-\infty}^{\infty} \frac{1}{N} \langle \boldsymbol{j}_T(\boldsymbol{q},t)\cdot\boldsymbol{j}_T(-\boldsymbol{q},0)\rangle e^{-i\omega t}dt, \quad (20)$$

$$C_L(\boldsymbol{q},\omega) = \int_{-\infty}^{\infty} \frac{1}{N} \langle \boldsymbol{j}_L(\boldsymbol{q},t)\cdot\boldsymbol{j}_L(-\boldsymbol{q},0)\rangle e^{-i\omega t}dt. \quad (21)$$

We adopted the DYNASOR package [45] to calculate the current correlation functions based on the MD trajectories, which can fully incorporate anharmonic effects and enable the exploration of temperature-dependent dispersion. If there are significant maxima in both $C_T(\boldsymbol{q},\omega)$ and $C_L(\boldsymbol{q},\omega)$, it signifies the discrete points in the $(\boldsymbol{q},\omega)$ plane on the dispersion curves [46].

Fig. 6 compares the transverse and longitudinal current correlation functions in the $(\boldsymbol{q},\omega)$ plane at different temperatures (see Fig. S12 for 200 and 500 K), corresponding to TA and LA collective vibrations [44–46], respectively. At the evaluated temperatures, the LA and TA vibrations exhibit clear dispersion relations below the Ioffe-Regel crossover frequency (see Ref. [44, 47, 48] for the Ioffe-Regel criterion delineated whether a phonon can be well-defined or not), suggesting that the phonons are well-defined excitations, potentially due to the short-range and medium-range ordering in a-SiO$_2$. Beyond the Ioffe-Regel crossover frequency, however, explicit lattice dispersion cannot be observed, and the peaks of $C_T(\boldsymbol{q},\omega)$ and $C_L(\boldsymbol{q},\omega)$ display pronounced broadening and softening with increasing $\boldsymbol{q}$ values (as depicted in Fig. 7), illustrating a liquid-like overdamped oscillation attenuation behavior [40–42, 49, 50]. Above the Ioffe-Regel crossover frequency, it is difficult to define the group velocity and phonon MFP for TA and LA, indicating the characteristics of liquid-like diffusion [50].

Additionally, as shown in Fig. 7, with decreasing temperature, the peak values of $C_T(\boldsymbol{q},\omega)$ and $C_L(\boldsymbol{q},\omega)$ at different $\boldsymbol{q}$-points gradually diminish, demonstrating the freezing of phonons as excitations at low temperatures. This phenomenon has been observed experimentally [89]. At 40 K, the region of TA and LA vibrations above the Ioffe-Regel crossover frequency nearly disappears, indicating a strong suppression of liquid-like diffusion characteristics at low temperatures. Hence, we expect phonons to be the primary thermal excitations at low temperatures, leading to the increase in the thermal conductivities of a-SiO$_2$ with rising temperatures. However, as the temperature rises, more phonons are excited (as evident in Fig. 7, the increasing peaks of $C_T(\boldsymbol{q},\omega)$ and $C_L(\boldsymbol{q},\omega)$ with temperature), and the intense phonon scattering events tend to reduce the thermal conductivity of a-SiO$_2$, similar to the behavior of crystalline materials at high temperatures. In Fig. 6, the Ioffe-Regel crossover frequencies of TA and LA become higher at high-temperature cases, and the domain resembling liquid-like diffusion becomes more pronounced. This suggests that phonon scattering and liquid-like diffusion [47, 49–51] collaboratively contribute to heat conduction, resulting in the temperature-insensitive thermal conductivity observed in a-SiO$_2$ at high-temperature regions.

Notably, Beltukov et al. [47] and Moon et al. [51] suggested that even below the phase transition temperatures, liquid-like diffusion in amorphous materials may be involved in the contribution to the total thermal con-ductivity due to atomic diffusion. Simultaneously, experimental evidence has been discovered supporting the notion of liquid-like thermal conduction in many complex crystals [42, 49, 50]. This work does not provide conclusive evidence for the contribution of atomic diffusion of liquid-like components to heat conduction at high temperatures, which could be a topic for our future work.

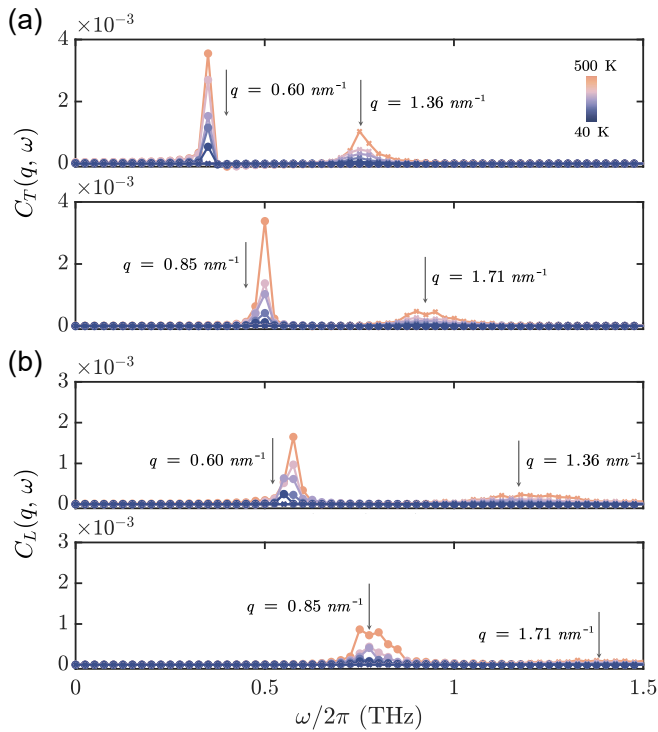

FIG. 7. (a) Transverse and (b) longitudinal current correlation functions as a function of $\omega/2\pi$ at different $\boldsymbol{q}$ points, including $\boldsymbol{q} =$ 0.6, 0.85, 1.36, and 1.71 nm$^{-1}$. The peaks of the current correlation function at various $\boldsymbol{q}$ points correspond to the blue region in Fig. 6, revealing a distinct dispersion relation below the Ioffe-Regel crossover frequency [44, 47, 48].

## V. SUMMARY AND CONCLUSIONS

In summary, by employing extensive MD simulations, we have explored the thermal transport properties of a-SiO$_2$ at various temperatures under the auspices of the accurate and efficient NEP model. Using a sophisticated melt-quench-anneal process with a slow quenching rate of $10^{11}$ K s$^{-1}$, the a-SiO$_2$ sample close to the experiment was generated in a system containing 73,728 atoms, validated through characterizations of pair-correlation function $g(r)$ for short-range order and static structure factor $S(q)$ for medium-range order. Based on the HNEMD method, we computed the thermal conductivity of a-SiO$_2$ from 20 to 2000 K and observed considerable deviations from experimental ones at low temperatures due to the absence of quantum statistics. Through the integration of spectral decomposition techniques and quantum statistics to correct classical thermal conductivities, we achieved strong consistency with experimental outcomes for bulk and 190 nm-thick film a-SiO$_2$ samples, thus affirming the significance of quantum statistical effects in predicting the thermal conductivity of amorphous materials.

Furthermore, we attempt to introduce the notion of collective excitations in liquids [40] or liquid-like matters [41, 42] to comprehend the thermal vibrations of a-SiO$_2$ at various temperatures. Employing atom velocities and trajectories derived from high-precision MD simulations, we computed the current correlation functions corresponding to TA and LA vibrations. The results indicate a distinct dispersion in TA and LA vibrations of a-SiO$_2$ at various temperatures below the Ioffe-Regel crossover frequency [44, 47, 48]. Hence, our expectation is that phonons, as excitations, can persist within a-SiO$_2$ and predominantly contribute to thermal conductivity at lower temperatures, consequently leading to the observed elevation of thermal conductivity with rising temperatures. As the temperature increases, however, more phonons are excited, and the effect associated with liquid-like diffusion becomes increasingly prominent. Therefore, we attribute the temperature-independent thermal conductivity in the high-temperature region of a-SiO$_2$ to the simultaneous engagement of excited intense phonon scattering and liquid-like diffusion in heat conduction. This work provides physical insights into unraveling the thermal transport properties of amorphous materials using MLP.

## ACKNOWLEDGMENTS

T.L. sincerely thanks Dr. Yanzhou Wang for providing constructive discussions. T.L. sincerely thanks for the Postgraduate Studentship from The Chinese University of Hong Kong. T.L., K.X., and J.X. acknowledge the support from the National Key R&D Project from Ministry of Science and Technology of China (Grant No. 2022YFA1203100), the Research Grants Council of Hong Kong (Grant No. AoE/P-701/20) and RGC GRF (No. 14220022). P.Y. is supported by the Israel Academy of Sciences and Humanities & Council for Higher Education Excellence Fellowship Program for International Postdoctoral Researchers. Z.Y. and J.X. acknowledge the support from the National Natural Science Foundation of China (No. 52106119, U20A20301).

**Data availability:**
Complete input and output files for the NEP training and testing are freely available at https://gitlab.com/brucefan1983/nep-data. Representative input and output files for different calculations in this work are freely available at https://github.com/Tingliangstu/Paper_Projects.

**Code availability:**

The source code and documentation for GPUMD are available at https://github.com/brucefan1983/GPUMD and https://gpumd.org, respectively. The source code and documentation for DYNASOR are available at https://gitlab.com/materials-modeling/dynasor and https://dynasor.materialsmodeling.org/, respectively.

**Declaration of competing interest:**

The authors declare that they have no competing interests.

# Supplemental Material:

# Mechanisms of temperature-dependent thermal transport in amorphous silica from machine-learning molecular dynamics

Ting Liang[1], Penghua Ying[2], Ke Xu[1], Zhenqiang Ye[3], Chao Ling[4], Zheyong Fan[*5], and Jianbin Xu[†1]

[1]*Department of Electronic Engineering and Materials Science and Technology Research Center, The Chinese University of Hong Kong, Shatin, N.T., Hong Kong SAR, 999077, P. R. China*
[2]*Department of Physical Chemistry, School of Chemistry, Tel Aviv University, Tel Aviv, 6997801, Israel*
[3]*College of Materials Science and Engineering, Shenzhen University, Shenzhen 518055, China*
[4]*School of Science, Harbin Institute of Technology, Shenzhen, 518055, P. R. China*
[5]*College of Physical Science and Technology, Bohai University, Jinzhou 121013, P. R. China*

# Contents

---

[*]Email: brucenju@gmail.com
[†]Email: jbxu@ee.cuhk.edu.hk

# Supplemental Notes

## Supplemental Note S1: Hyperparameters for NEP training

We used GPUMD-v3.5 to train the NEP model, which is a NEP4 model. The `nep.in` input file for GPUMD reads:

```
version      4          # NEP4
type         2 O Si     # Two elements: oxygen (O) and silicon (Si)
cutoff       6 4        # Radial and angular cutoffs
n_max        8 8        # Size of radial and angular basis
basis_size   8 8        # Number of radial and angular basis functions
l_max        4 2 0      # Expansion order for angular terms
neuron       80         # Number of neurons in the hidden layer
lambda_1     0.1        # Weight of L1 regularization term
lambda_2     0.1        # Weight of L2 regularization term
lambda_e     1.0        # Weight of energy loss term
lambda_f     1.0        # Weight of force loss term
lambda_v     0.1        # Weight of virial loss term
population   50         # Population size used in SNES algorithm
batch        10000      # Batch size for training (Full batch)
generation   1000000    # Number of generations used by SNES algorithm
```

The cutoff radii for radial and angular descriptor parts are $r_\mathrm{c}^\mathrm{R} = 6$ Å and $r_\mathrm{c}^\mathrm{A} = 4$ Å, respectively. For the radial descriptor components, we use 9 (`n_max + 1`) radial functions, each one being a linear combination of 9 (`basis_size + 1`) basis functions. For the angular descriptor components, we also use 9 radial functions linearly combined from 9 basis functions (Chebyshev polynomials). Following the atomic cluster expansion (ACE) approach [1] for constructing the angular descriptor components, we use three- and four-body correlations in the spherical harmonics up to degree $l = 4$ and $l = 2$, respectively. The total descriptor vector for one element thus has $9 + 9 \times 5 = 54$ components. With 80 neurons in a single hidden layer, the neural-network architecture for each element can be written as 54-80-1. With $9 \times 9 + 9 \times 9 = 162$ trainable descriptor parameters for each pair of elements, the total number of trainable parameters for our NEP model of silica is $(54 + 2) \times 80 \times 2 + 1 + 162 \times 2^2 = 9609$. The final NEP model was trained for one million generations (steps) with a full batch size.

# Supplemental Figures

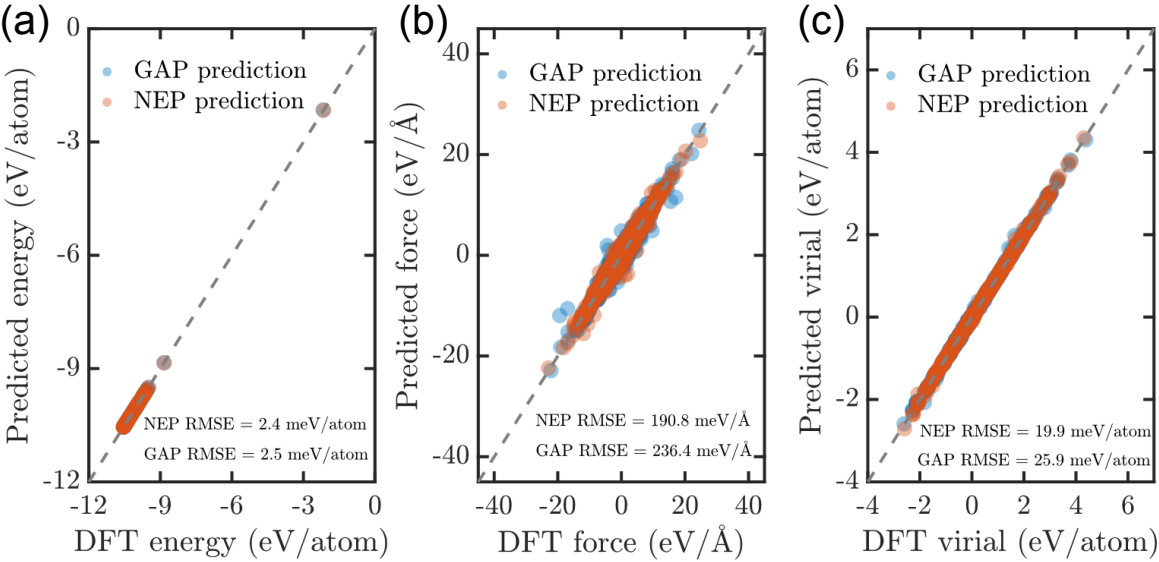

Figure S1: Parity plots of (a) energy, (b) force, and (c) virial for the NEP model trained in this work and the GAP model taken from Ref. [2] for the 261 structures we selected as the test set in NEP training. The NEP model was trained using 2348 structures excluding the 261 structures here, while the GAP model was trained using 3074 structures including the 261 structures here.

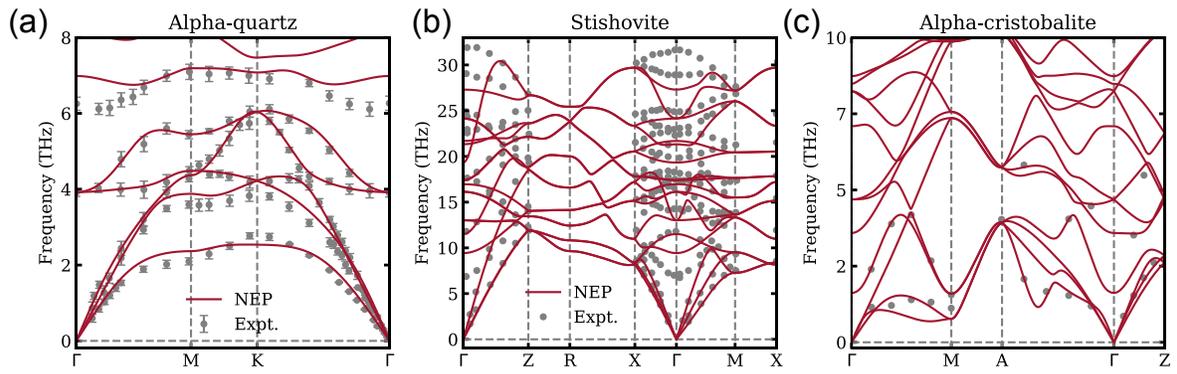

Figure S2: The comparison of NEP-calculated phonon dispersion for different silica crystal structures with experiment. Experimental results in (a) alpha-quartz, (b) stishovite, and (c) alpha-cristobalite are taken from [3], [4], and [5], respectively.

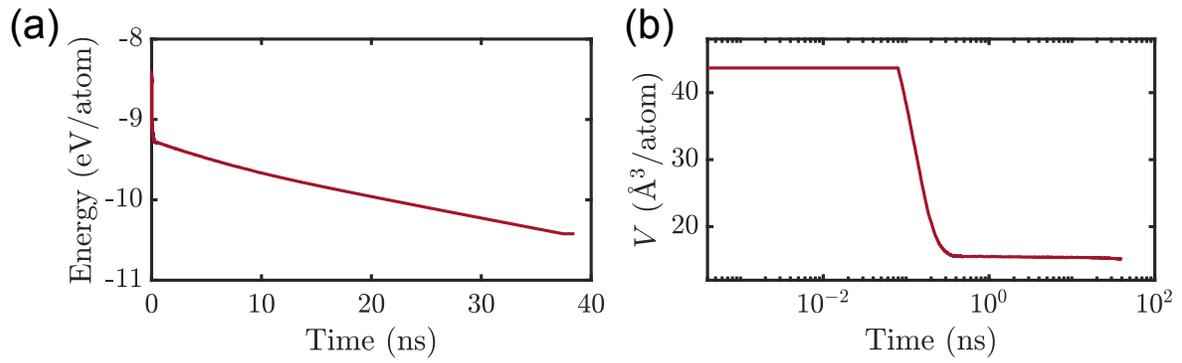

Figure S3: The evolution of (a) the energy and (b) the cell volume, $V$ during the melt-quench-anneal process. The slow quench rate rate of $10^{11}$ K s$^{-1}$ enabled the simulated system transition from a low-density liquid to a high-density amorphous state.

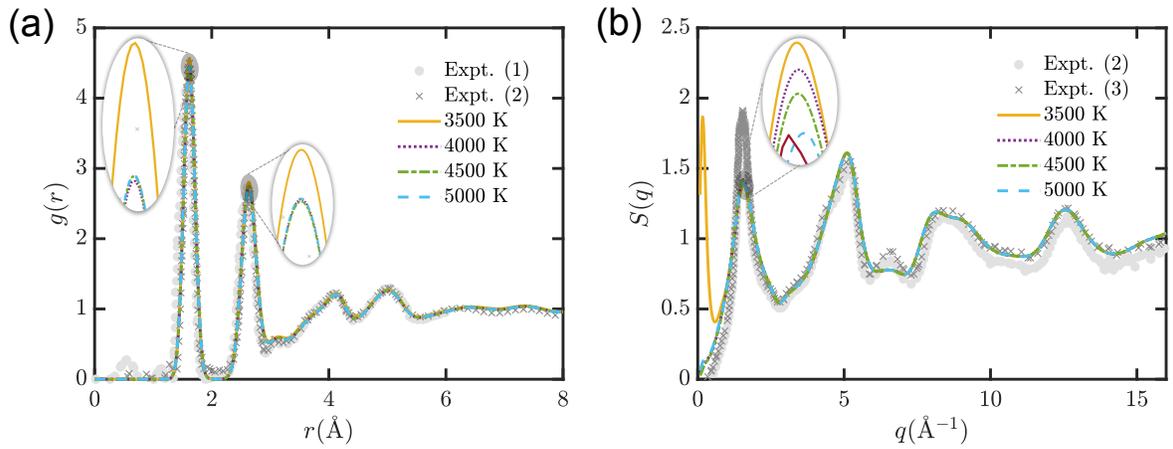

Figure S4: The effect of different starting temperatures of the quenching process (corresponding to point **B** in Figure 2 of the main text) on (a) pair-correlation function $g(r)$ and (b) X-ray structure factor $S(q)$. The simulated tests are performed on a system of 73,728 atoms at a quenching rate of $10^{12}$ K s$^{-1}$. The starting temperature of the quenching process above 4000 K has no significant effect on $g(r)$, while $S(q)$ at 4000 K is more comparable to the experimental results ("1"; ref. [6], "2"; ref. [7], and "3"; ref. [8]). Consequently, all our a-SiO$_2$ samples are generated at a starting temperature of 4000 K during the quenching process.

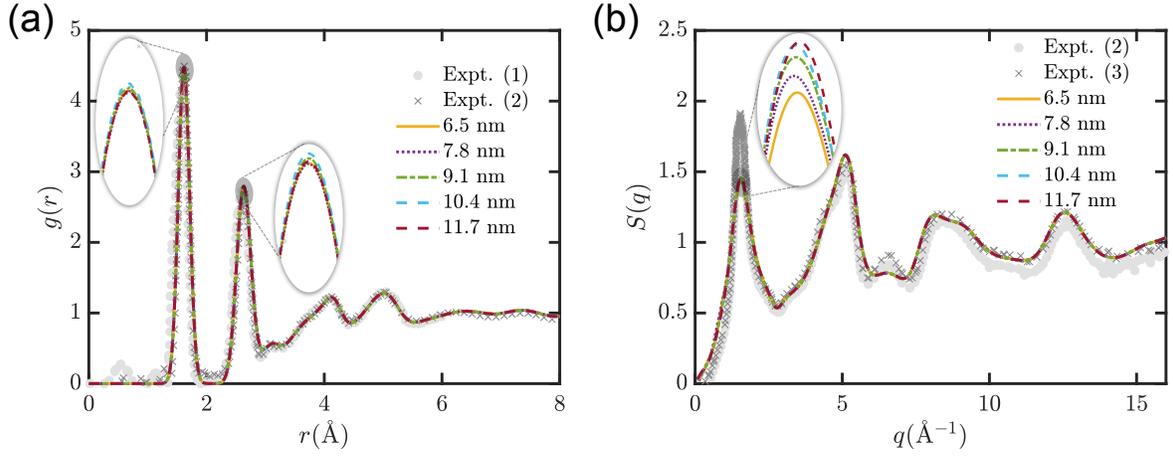

Figure S5: The comparison of (a) the $g(r)$ and (b) the $S(q)$ calculated for different system sizes of a-SiO$_2$ (with a quenching rate of $10^{11}$ K s$^{-1}$) with experiments. Labeled as the lengths of the final a-SiO$_2$ structures obtained after the melt-quench-anneal process, and the 6.5, 7.8, 9.1, 10.4, and 11.7 nm systems contain 18,000, 3,1104, 49,392, 73,728, and 104,976 atoms, respectively. The first peak of the $g(r)$ (reflecting short-range order) remains unchanged with an increase in system size. However, the first sharp diffraction peak of the $S(q)$ (reflecting medium-range order) rose slightly with increasing system size. After considering the computational efficiency, the system containing 73,728 atoms is ultimately chosen for our subsequent calculations on the thermal conductivity of a-SiO$_2$.

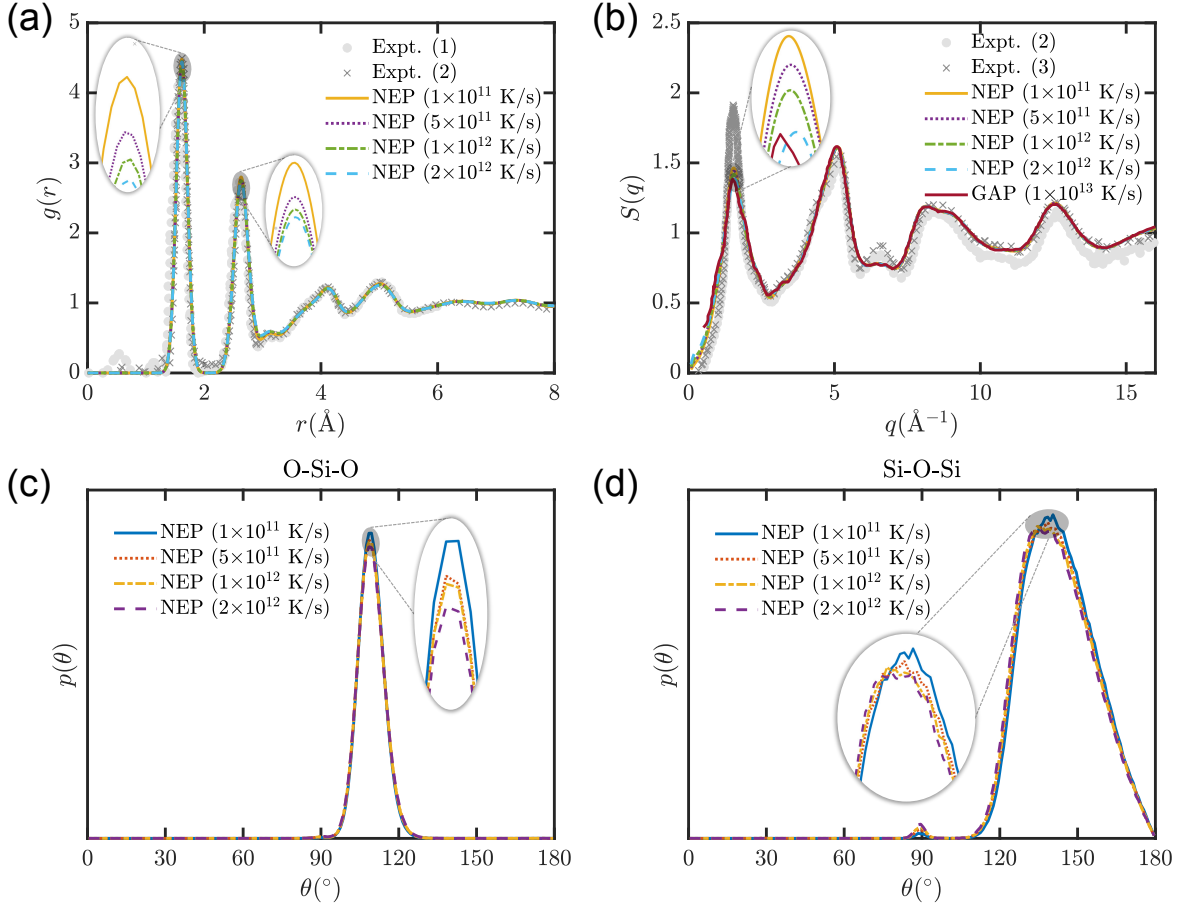

Figure S6: (a, b) The comparison with experiments of $g(r)$ and $S(q)$ obtained at 300 K in a system of 73,728 atoms on different quenching rates. The $S(q)$ data obtained from the GAP model calculations are extracted from reference [9]. (c, d) The comparison of the angular distribution function $p(\theta)$ at different quenching rates, where (c) corresponds to the O-Si-O bond and (d) corresponds to the Si-O-Si bond. Overall, a lower quenching rate produces samples that are closer to experiments. Specifically, a lower quenching rate makes the first diffraction peak of $S(q)$ closer to the experimental peak. Consequently, the thermal conductivity is calculated by using the a-$SiO_2$ samples with a quenching rate of $10^{11}$ K s$^{-1}$.

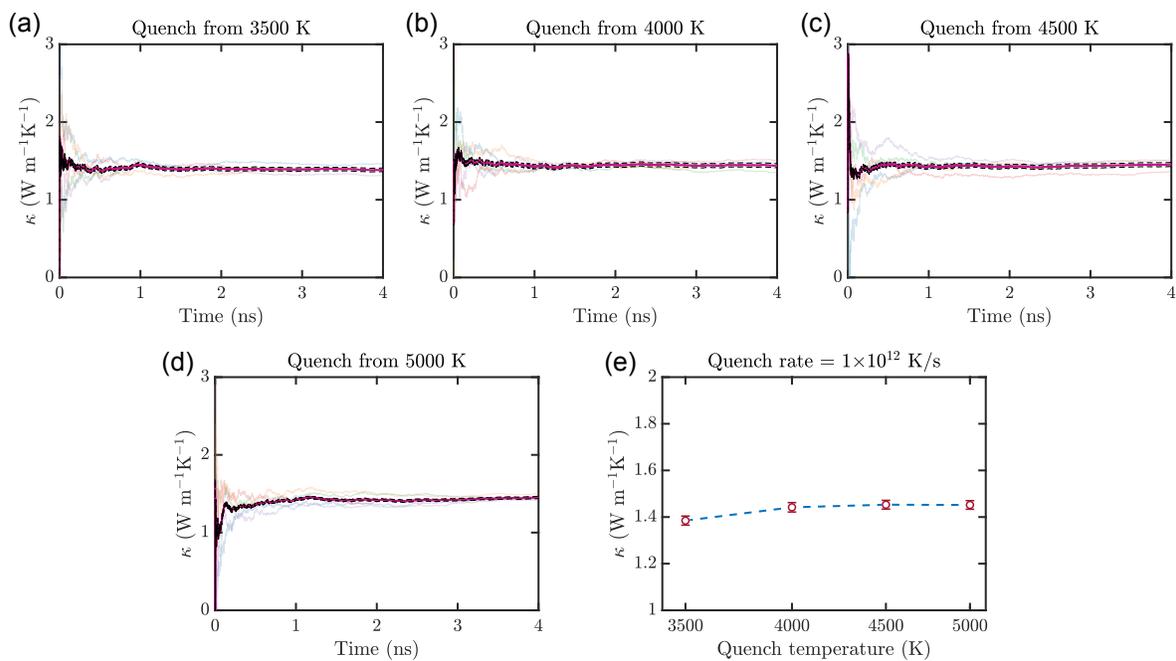

Figure S7: (a-d) Cumulative average of the thermal conductivity $\kappa$ for different starting temperatures of the quenching process as a function of the HNEMD production time. In each panel, the thin transparent lines are from five independent runs, and the thick magenta solid and black dashed lines represent the average and error bounds from the individual runs. (e) Variations of thermal conductivity $\kappa$ calculated by HNEMD with different quenching onset temperatures. The $\kappa$ is calculated at zero pressure and 300 K after obtaining the final a-SiO$_2$ structure.

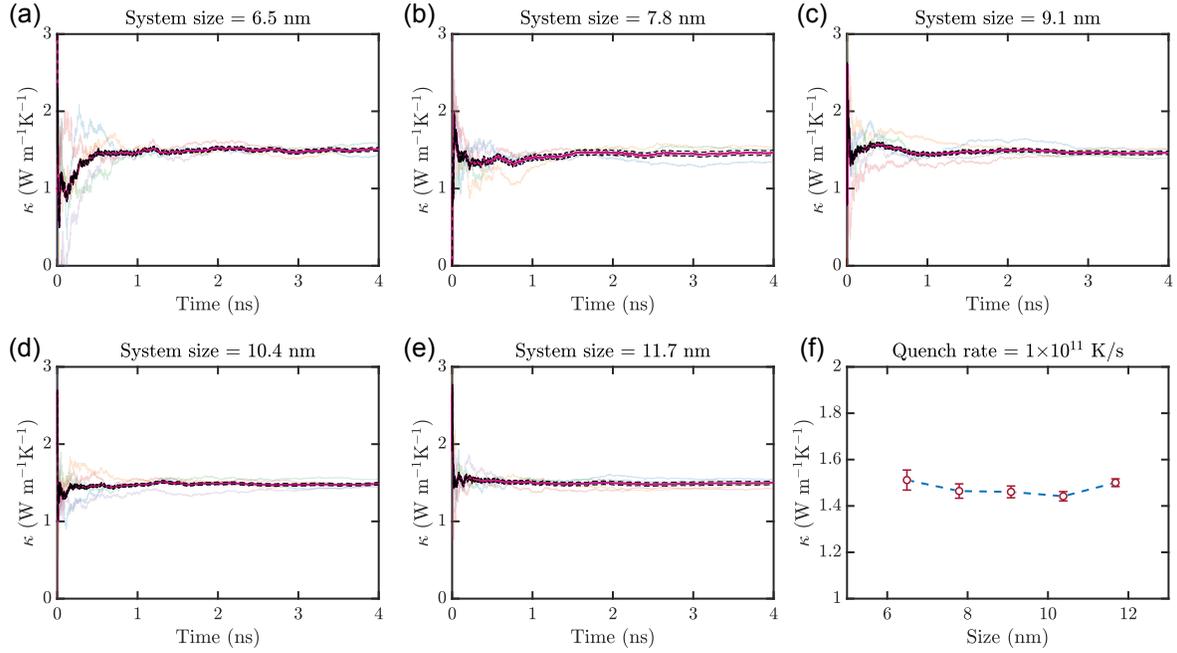

Figure S8: Cumulative average of the thermal conductivity $\kappa$ at 300 K and zero pressure for different system lengths as a function of the HNEMD production time. The side length of the cubic simulation cell from (a) to (e) increases from $L = 4.5$ nm to $L = 11.7$ nm. The meanings of the different line shapes are the same as in Fig. S7. (f) Variations of thermal conductivity $\kappa$ calculated by HNEMD with different system lengths.

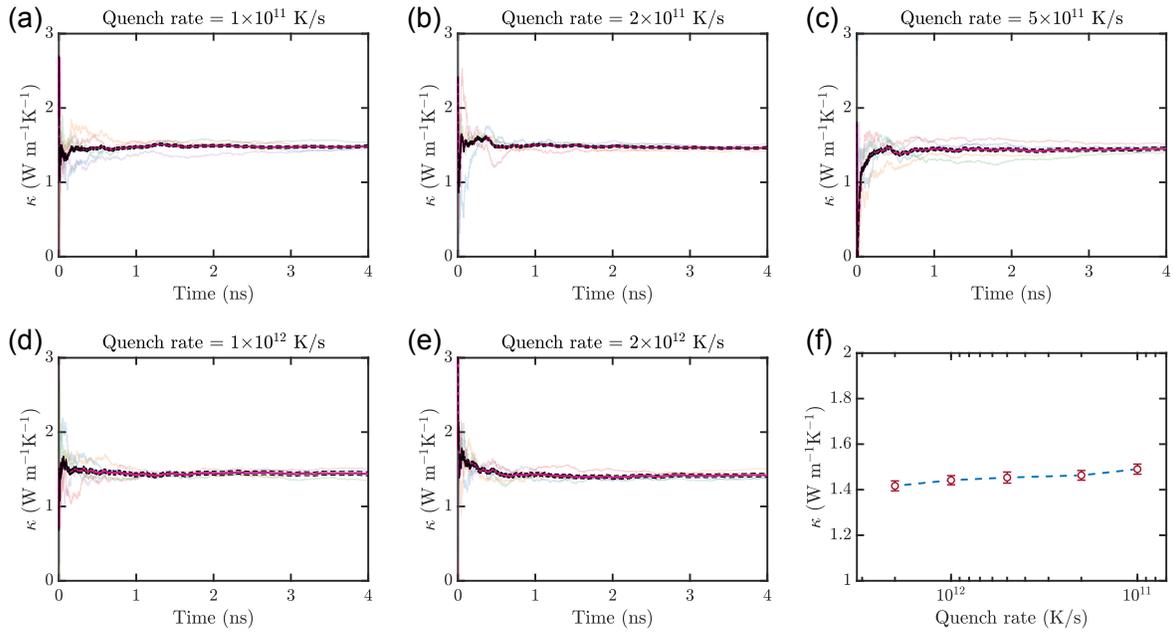

Figure S9: (a-e) Cumulative average of the thermal conductivity $\kappa$ at 300 K and zero pressure for different quenching rates as a function of the HNEMD production time. The meanings of the different line shapes are the same as in Fig. S7. (f) Variations of thermal conductivity $\kappa$ calculated by HNEMD with different quenching rates.

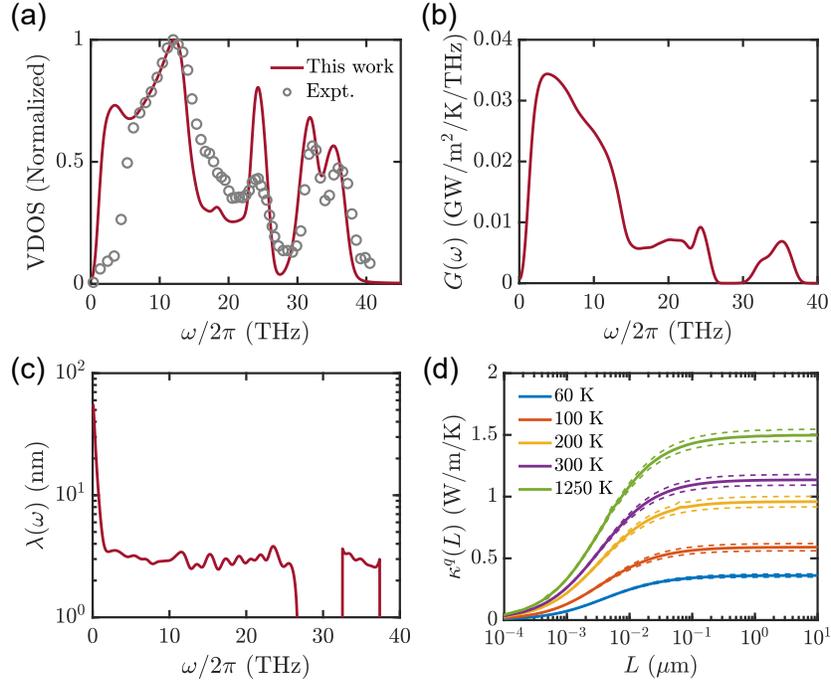

Figure S10: (a) A comparison between the normalized vibrational density of states (VDOS) at 300 K and experimental data [10]. (b) The classical spectral thermal conductance $G(\omega)$ at 20 K. (c) Phonon MFP $\lambda(\omega)$ as a function of phonon frequency $\omega/2\pi$ for a-SiO$_2$ at 300 K. (d) Quantum-corrected length-dependent thermal conductivity $\kappa^{\mathrm{q}}(L,T)$ of a-SiO$_2$ in the transport direction ($z$-direction). The dashed lines indicate the standard deviation derived from five independent MD simulations.

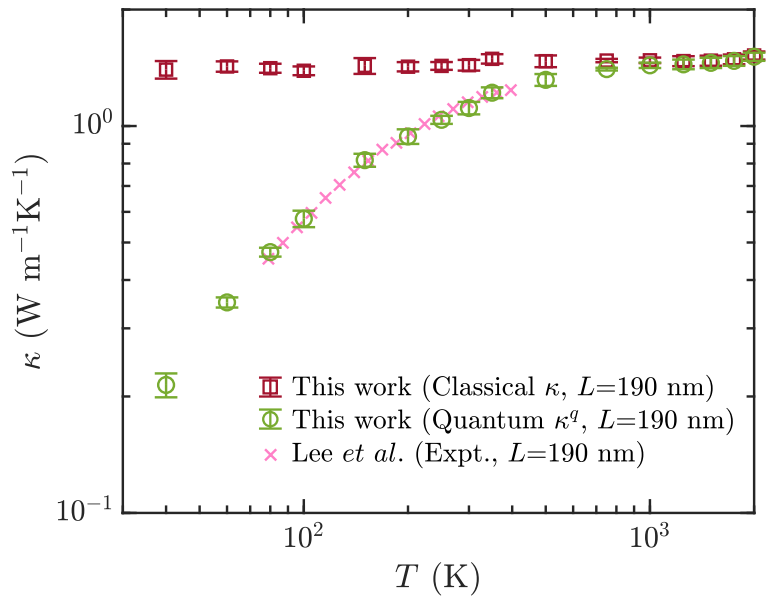

Figure S11: (a) Thermal conductivity $\kappa$ of a-SiO$_2$ samples with a thickness of $L = 190$ nm in the temperature range from 40 to 2000 K. Experimental data [11] is available within the temperature span of roughly 80 to 400 K.

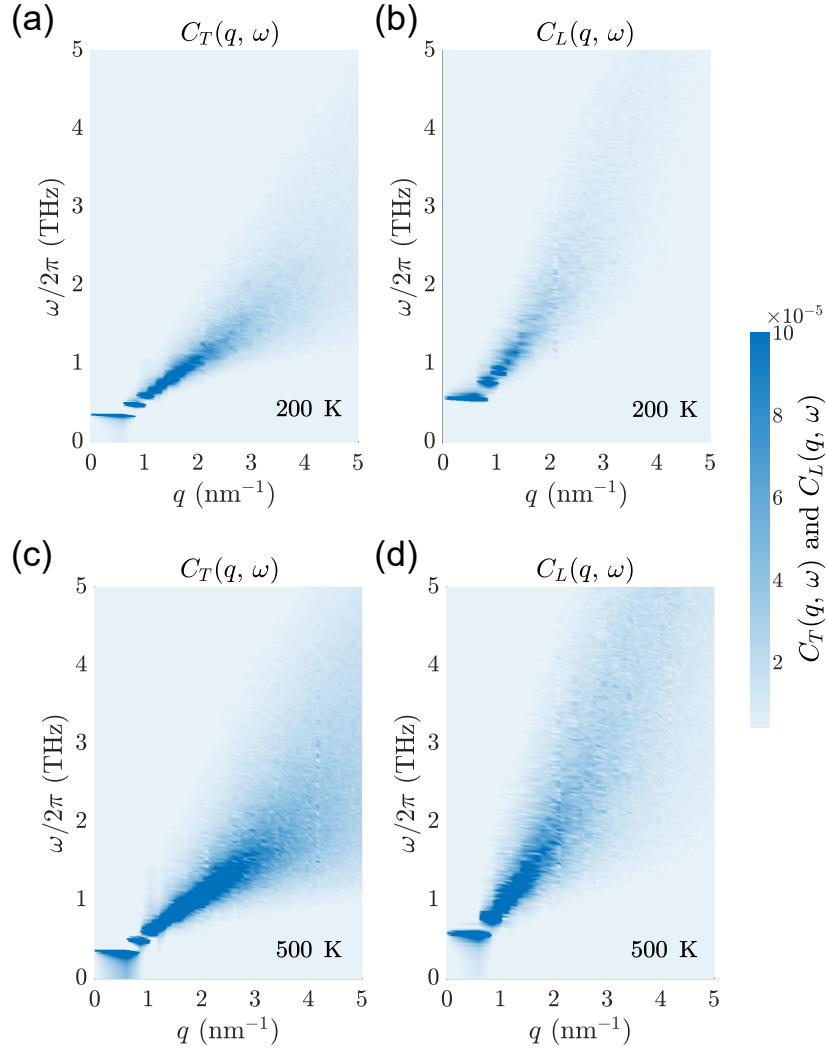

Figure S12: (a) Transverse and (b) longitudinal current correlation functions as a function of $\boldsymbol{q}$ vector and $\omega/2\pi$ at 200 K. (c) Transverse and (d) longitudinal current correlation functions as a function of $\boldsymbol{q}$ vector and $\omega/2\pi$ at 500 K.

# Supplemental References